\documentclass[10pt, oneside]{book}

\usepackage{amsfonts,amsmath,amssymb,amsthm, mathtools}

\usepackage{cite}
\usepackage[dvipsnames]{xcolor}

\usepackage[a4paper, margin=1in]{geometry}

\usepackage[T1]{fontenc}

\usepackage{graphicx}

\usepackage{caption, subcaption}

\usepackage{authblk}

\DeclareUnicodeCharacter{2009}{\,}

\usepackage{thmtools}

\renewcommand{\emph}[1]{\textit{#1}}

\usepackage{tikz} 
\usetikzlibrary{shapes.geometric, 
                arrows,
                decorations.pathmorphing,
                decorations.pathreplacing, 
                decorations.shapes,
                positioning,
                patterns,
                arrows.meta, 
                automata, 
                shapes.misc,
                calc,
                backgrounds,
                fit,
                }

\begin{document}

\title{White Paper on Quantum Internet Computer Science Research Challenges}



\author[1,2,3]{Thomas R. Beauchamp}
\author[1,2,3]{Scarlett Gauthier}
\author[1,2,3]{Stephanie Wehner}

\affil[1]{QuTech, Delft University of Technology, Delft, The Netherlands}
\affil[2]{Kavli Institute of Nanoscience, Delft University of Technology, Delft, The Netherlands}
\affil[3]{Quantum Computer Science, EEMCS, Delft University of Technology, Delft, The Netherlands}

\maketitle

\tableofcontents

\chapter{Executive Summary}

The aim of a quantum network is to enable the generation of end-to-end entangled links between end nodes of the network, so that they can execute quantum network applications. 
To facilitate this, it is desirable to have robust control of the network in order to be able to provide a reliable service to the end nodes.
In recent work \cite{journal-arx}, we proposed a modular control architecture for a generate-when-request type network.
This control architecture enables quantum network applications to be executed on end nodes running a modern operating system such as QNodeOS \cite{QNodeOS}.
In that work, we performed an evaluation of our architecture based on a proof-of-concept implementation.

In the course of performing this evaluation, we discovered many outstanding questions and challenges. 
These relate not only to implementing our specific control architecture, but also to the design of any quantum network control architecture.

Here, we describe some outstanding questions and challenges, discuss possible solutions, and identify where existing protocols require adaptation or new protocols must be designed.

\chapter{Introduction}
The realization of a quantum internet will enable the use of networked applications beyond what is currently possible with the classical internet. 
These include the ability to perform verifiably secure secret sharing \cite{ekert_quantum_1991}, secure remote computation \cite{bqc1}, and secure leader election \cite{tani_exact_2012}, among many others. 
The aim of a quantum network architecture therefore should be to provide a service to end nodes which ensures that these applications can be successfully executed. 
End nodes must be able to use the network's internal resources to generate end-to-end entangled links between themselves. This can be achieved in several ways, such as through a best-effort approach based on packet switching, analogous to the classical internet \cite{LeveragingInternetPrinciples}, or by constructing a network schedule that determines which nodes can use which resources and when. To satisfy a substantial proportion of demands for entanglement, the resource requirements of a best-effort architecture—in terms of quantum memory quality and quantity at each network link—scale with the number of demands. In contrast, a schedule-based architecture may have constant resource requirements when network scheduling is paired with effective admission control on arriving demands.

There are many questions which remain to be answered in the domain of quantum internet computer science research. 
Many of these overlap with questions that arise in quantum computer science and quantum computing in general, for example how to coherently control many qubits. 
However, there are many questions which are unique to quantum \textit{networking}. 
For example, how does one generate or distribute entanglement across vast distances. 
Here we will focus on some of these challenges and questions which pertain to controlling a quantum network. 

In order to motivate some of the questions we raise, we will use as reference the network control architecture we introduced in \cite{journal-arx}
This network architecture is compatible with modern operating systems for quantum end nodes, such as QNodeOS \cite{QNodeOS}. 
In particular, the architecture provides a method for computing network schedules and distributing them to nodes in the network.
This is mediated by a central controller, which periodically schedules time for entanglement generation in response to demands from (pairs of) end nodes. 
To evaluate this architecture, we created a proof-of-concept implementation and conducted numerical simulations.
This implementation used a network scheduler based on an adapted earliest-deadline-first network scheduler, equipped with simple admission control rules to try and prevent overloading the network. 
The results obtained from evaluation of our proof-of-concept implementation exposed questions and challenges which require addressing in order to ensure high quality service can be delivered to end-nodes. Quality of service refers to satisfaction of end-node or network defined performance metrics.

Here we move beyond the previously identified limitations of our implementation and raise questions highlighting areas where general quantum network control architectures require further development.
In addition to exposing the challenges we believe to be the most pressing, we also give some ideas for possible solutions. 

The rest of this manuscript is organized as follows: in Chapter~\ref{ch: general requirements}, we discuss general requirements for near-term quantum internet architectures; in Chapter~\ref{ch: arx}, we summarize key features of our proposed architecture, our proof-of-principle implementation, and evaluations; in Chapter~\ref{ch: future considerations}, we pose questions about the future development of quantum network control architectures and propose potential solutions; in Chapter~\ref{ch: conclusions}, we conclude and discuss our outlook on these challenges.

\chapter[General Requirements]{General Requirements for Near-Term Quantum Internet Architectures}
\label{ch: general requirements}
\section{Compatibility with modern end node operating systems}
A key requirement of any quantum network architecture is that it is compatible with modern operating systems for quantum end nodes. 
The major example is QNodeOS \cite{QNodeOS} and its upgrade Qoala \cite{vecht_qoala_2025}.
This system decomposes applications into node-specific programs, which are further divided into instruction blocks that can be scheduled for execution. This decomposition enables effective multi-tasking on a single end node, allowing multiple applications to run simultaneously.

One such block of instructions is a \verb|Request| routine for entanglement generation.
However, to effectively schedule these blocks, an end node requires a network schedule stating between which times entanglement generation with the partner node(s) may take place.
Therefore, any near-term network architecture should be able to provide for the creation and distribution of such a network schedule to the various quantum network nodes. Best-effort quantum network architectures that do not provide network schedules may face a major integration challenge with end node operating systems.

\section{Compatibility with modern hardware}
In the near-term, the capability of quantum network devices will be severely limited. 
For example in experiments on leading Nitrogen Vacancy (NV) center hardware, Stolk and van der Enden \textit{et al.} in \cite{stolk_metropolitan-scale_2024} reported memory lifetimes of eleven milliseconds and end-to-end entanglement generation rates of one link every forty seconds over a net distance of fifteen kilometers. 
Similarly in experiments on trapped-ion hardware, Krutyanksi \textit{et al.} in \cite{krutyanskiy_telecom-wavelength_2023}, reported memory lifetimes of sixty-two milliseconds. These impressive demonstrations indicate progress in the development of useful quantum network nodes. However, the demonstrated memory lifetimes and rates of entanglement availability pose limitations on the implementation of quantum network architectures. 

Firstly, the limited nature of modern quantum memories greatly restricts the ability to perform, e.g., entanglement buffering. 
Given that the average time to create a new entangled link even between neighboring nodes is many times larger than the memory lifetimes on those nodes,  the likelihood of having more than a single elementary link between a pair of nodes in the network is very low.
Therefore, architectures for near-term networks should assume that the network operates under a \textit{generate-when-requested} model, where it is assumed no entanglement is stored in the network, as opposed to a pre-loaded network, where it is assumed that entanglement is very prevalent in the network. 

Secondly, classical communication times can become a significant limiting factor in performance. 
For example, sending even a simple classical message over a TCP/IP connection over a distance of ten kilometers can take from fifty microseconds to a few milliseconds. 
Over hundreds of kilometers, these times can increase to tens or even hundreds of milliseconds.
This is already a significant portion of the memory lifetime of any given node. 
Therefore, if classical communication is required after every elementary entangled link is generated it will strongly limit the ability of the network to produce end-to-end entanglement. 
This is especially true in network architectures such as \cite{cicconetti_request_2021}, where routing decisions are made each time entanglement is generated, thereby introducing delays due to both classical computation times and communication.  
Near-term quantum network architectures need to stringently account for the impact of classical communication requirements.

\section{Compatibility with quantum network applications}
Fundamentally, the aim of a quantum internet is to allow the execution of quantum network applications. 
Quantum network applications can be classified into three categories: \textit{Measure-Directly} (MD), \textit{Create-and-Keep} (CK), and \textit{High Availability} (HA).
In MD applications, as soon as an entangled link is created it is measured and consumed.
The prototypical example of such an application is QKD \cite{ekert_quantum_1991}. 
In order to gain the full quantum advantage of unconditionally secure cryptographic keys, it is required to use end-to-end entangled links. 
However, modern QKD networks either only span small enough distances for direct entanglement generation to be possible (e.g. \cite{dynes_cambridge_2019}), or have to use trusted intermediary nodes (e.g. \cite{martin_madqci_2024}). 
Applications which are members of the CK class require many co-existing entangled links to be present in memory before an instance can be executed. 
Here, the prototypical example is \textit{Blind Quantum Computing} (BQC)  \cite{bqc1, bqc2, childs_secure_2005},
 where many states need to be teleported from the client to the server performing the calculation before the calculation can begin.
In HA applications the key requirement is that entangled links are highly available to satisfy consumption requests, which may arrive irregularly. Here, the prototypical example is load balancing of a classical network \cite{loadBalancing, HFT}, where an entangled pair can be consumed in determining to which of two servers to route traffic.

The implementation of an application can affect its class. 
For instance, an application in the MD class implemented with distillation of end-to-end entanglement is actually in the CK class in this context.

\chapter{Our Proposed Architecture}
\begin{figure*}[t]
    \centering

    \begin{tikzpicture}[
            node distance = 1cm,
            action/.style={rectangle,
                            draw=black!60,
                            fill=white,
                            very thick,
                            align=center,
                            minimum width=2cm,
                            minimum height = 1.5cm,
                            },
            decision/.style={diamond,
                            draw=black!60,
                            fill=white,
                            very thick,
                            align=center,
                            minimum width=2cm,
                            minimum height = 2cm,
                            aspect=2,
                            },
            A/.style={->,thick}
]

    \node[action] (NCU) {Network\\Capability\\Update};
    \node[action] (CN) [right=of NCU] {Capability\\Negotiation};
    \node[action] (DS) [right=of CN] {Demand\\Submission};
    \node[action] (NS) [right=of DS] {Network\\Scheduling};
    \node[action] (NdS) [right=of NS] {Schedule\\Dsitribution};

    \draw[A] (NCU) -- (CN);
    \draw[A] (CN) -- (DS);
    \draw[A] (DS) -- (NS);
    \draw[A] (NS) -- (NdS);

\end{tikzpicture}

\caption[Flow of information in the architecture]{Flow of information through the architecture proposed in \cite{journal-arx}. The processes of `Network Capability Update' and `Capability Negotiation' allow the nodes to gather enough information to be able to submit a unified demand. These demands are then used to construct a central network schedule, which in turn is used when the operating systems of nodes construct their local schedules. Reproduced from \cite{journal-arx}.}
    \label{fig:Information Flow}
\end{figure*}
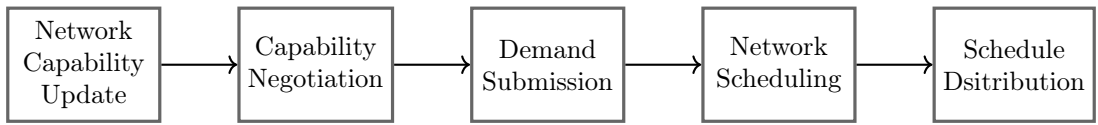
\label{ch: arx}
For convenience, we give a brief overview of the architecture we proposed in \cite{journal-arx}. A summary of the flow of information through our architecture can be seen in Figure~\ref{fig:Information Flow}.

We model the quantum network as being a \textit{centrally controlled}, \textit{Generate-When-Requested} (GWR) network. 
This means firstly that there exists a central controller which has jurisdiction over determining when entanglement generation can occur and between which pairs of nodes. 
This is achieved by computing fixed-length network schedules and distributing them ahead of time to the network components, which can then execute them at the appropriate times. 
Secondly, the GWR aspect of the network means that we do not assume that any entanglement is stored in the network other than by end nodes, and that end-to-end link generation only occurs as a response to the submission of a demand by a pair of end nodes. 
This is in contrast to what we refer to as \textit{pre-loaded} networks (e.g. \cite{cicconetti_request_2021, ESDI}) where end-to-end entangled links are constantly being generated and therefore are immediately available to an application to use. 

Our architecture consists of three main sets of activities: 1. demand submission; 2. admission control and network scheduling; and 3. schedule distribution. 
During demand submission, pairs of end nodes obtain information about the capabilities of the network, come to a common understanding about how the application is to be executed and submit a single unified demand for service to the central controller. Demands are requests for a minimum number of \textit{packets of entanglement}, to be generated before an expiry time, with a target rate $R$ of packet generation over the entire interval of time in which the demand receives service.
These demands are then ingested by the central controller, which periodically applies an admission control protocol to all pending demands, determining whether they can be accepted for network scheduling or not. 
Once the set of demands to be scheduled is confirmed, a schedule is computed and finally distributed to the various components. 
Admission control ensures that the network doesn't become significantly overloaded and additionally, that network schedules can be computed in a timely manner. 

The network schedules themselves consist of a number of finite-length blocks of time called \textit{packet generation attempts} (PGAs).
During each packet generation attempt, resources along a path in the network are reserved solely for the use of generating entanglement between a specific pair of nodes in response to an accepted demand. 
So long as the required resources are available, there may be many PGAs scheduled at the same time, allowing multiple demands to be served at once.

\section{Evaluation results}

\begin{figure}
    \centering
    \includegraphics[]{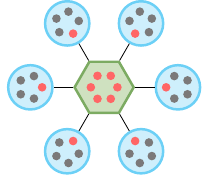}
    \caption[Example of a star-topology network with 6 nodes.]{Example of a star-topology network with 6 nodes. The outer circles represent the end nodes and the central hexagon a central junction node. 
    The orange dots represent communication qubits and the black dots represent memory qubits.
    The central node is capable of creating an entangled link with each of the end nodes and performing entanglement swaps to create end-to-end links between pairs of end nodes.
    Reproduced from \cite{journal-arx}.}
    \label{fig:implementation star network}
\end{figure}

\begin{figure}
    \begin{subfigure}[b]{0.48\textwidth}
        \includegraphics[width=\linewidth]{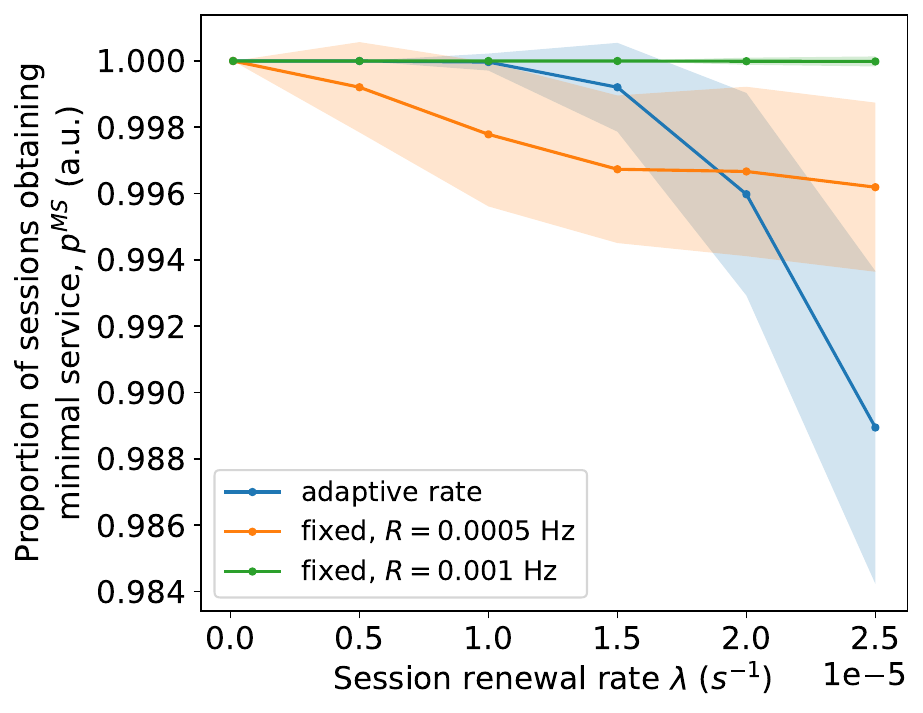}
        \caption{}
    \end{subfigure}
    \begin{subfigure}[b]{0.48\textwidth}
        \includegraphics[width=\linewidth]{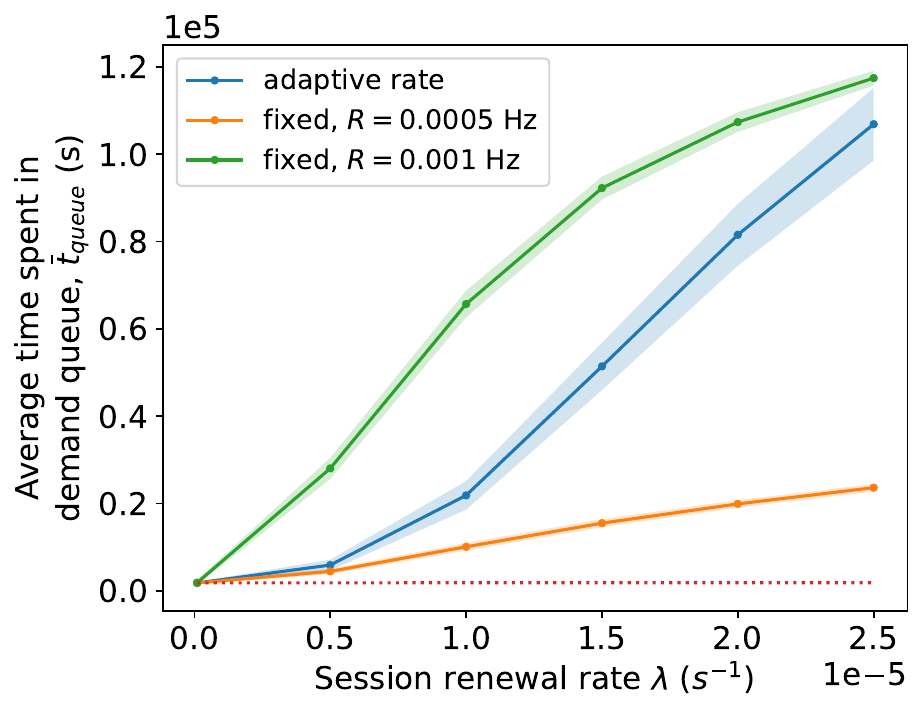}
        \caption{}
    \end{subfigure}
    
    \caption{
        Results from simulations for client-server CKA on a six-node star network.
        (a) Proportion of initiated sessions which obtained minimal service from the network.
        (b) The average time a demand spent in the demand queue. The shaded region represents $\pm1\sigma$, where $\sigma$ is the standard deviation.
        The red dotted line represents the expected value of $t_{\text{queue}}$ for a single pair of end nodes submitting demands.
        The total simulated time was 360 days.
        Reproduced from \cite{journal-arx}.
    }   
    \label{fig:arx-paper-cka-results}
\end{figure}

Here we summarize the results obtained from our evaluation of the network architecture. Details on the admission control and scheduling protocols with which our architecture was equipped in these evaluations can be found in Chapter~\ref{ch: future considerations}, where they are contrasted against proposals for improved protocols.

The evaluation was performed on the star-topology network illustrated in Figure~\ref{fig:implementation star network}, which consists of six end nodes connected to a central hub.  
The central hub is known as a junction node. It has several communication qubits that can be used to generate entangled links with end nodes, and it can perform entanglement swaps to create end-to-end links between pairs of end nodes. 
In our evaluations pairs of end nodes submitted requests to execute a CK type test application based on BQC (hereafter CKA). 
One node was nominated as the `server', and the other five `client' nodes submitted requests to do BQC with the server. 
Demands were marked as complete and subsequently terminated by the central controller as soon as one of the following conditions was met: \begin{enumerate}
    \item All packets of entanglement required to satisfy the demand were successfully generated, an event we refer to as the demand \textit{obtaining minimal service}, or
    \item The demand expiry time was reached.
\end{enumerate}

Whenever a demand from a pair of end nodes was marked complete, either due to obtaining minimal service or due to demand expiry, a new demand was submitted at a time $\mathrm{Exponential}(\lambda)$ later. The parameter $\lambda$ is therefore a session renewal rate parameter, determining the frequency with which new demands arrived at the central controller. In total, each pair of end nodes contributed to the service load on the central controller via the session renewal rate $\lambda$ as well as the rate of packet generation $R$ requested by each demand.

The results of these evaluations are included in Figure~\ref{fig:arx-paper-cka-results}. We evaluate the response of the central controller to load in three different scenarios: In the first two, demands request a fixed rate of packet generation $R$ of either $0.001$ Hz or $0.0005$ Hz; in the third regime, demands request an adaptive rate, allowing the central controller to dynamically adjust the service rate to prioritize delivering minimal service under the instantaneously observed load conditions. In each scenario we studied the response to increasing service load by repeating simulations with gradually increasing session renewal rates, $\lambda$. We measure the response to service load by the proportion of sessions (demands) obtaining minimum service, as well as the average time a demand spent in a queue before receiving service (a response time). 
In all scenarios investigated, over 98\% of demands were able to obtain minimal service. 
However, it is clear that the proportion of demands which obtain minimal service depends on both the session renewal rate as well as on the rate of packet generation attempts $R$ which is requested. In particular, in both the fixed rate $R=0.0005$ Hz scenario and in the adaptive rate scenario, the proportion of demands obtaining minimal service decreased as the load was increased, by increasing $\lambda$. 
In all scenarios, we observed significant demand queuing times (response times), which increased significantly as the session renewal rate $\lambda$ was increased. 
We also performed simulations with a measure-directly test application, based on QKD, and obtained similar results as for the CKA application. 

These results indicate that the admission control and scheduling protocols with which our network architecture were equipped are not robust against a high service load or increases in service load on the network. 
A takeaway is that the architecture needs to be equipped with new protocols to ensure that as the load on the network increases the service which is provided to users doesn't degrade.

\chapter[Future Quantum Network Control Architectures]{Considerations for Future Implementations of Quantum Network Control Architectures}\label{ch: future considerations}
\section{Good admission control}
In the evaluations we performed in \cite{journal-arx}, we employed a simple admission control routine which checked two conditions: 
\begin{enumerate}
  \item Does the utilisation of any resource, defined by $$U_r = \sum_{\tau:r\in\rho_\tau} E_\tau R^\texttt{attempt}_\tau,$$ exceed a certain limit, where $E_{\tau}$ is the execution time of a PGA for demand $\tau$ requiring resource $r$, and $R^{\texttt{attempt}}_{\tau}$ is the required rate of PGA scheduling for that same demand;
  \item Does the expected time to compute the network schedule exceed the cut-off required to distribute the schedule on time.

\end{enumerate}
We also implemented a first-in-first-out (FIFO) demand queue, into which demands were placed once they passed admission control. 
Demands were removed from the queue either by being accepted for scheduling or else by expiring or being terminated.

We observed that this was insufficient to ensure that pairs of end nodes were able to generate the minimum number of packets requested before the demand expired. 
Indeed, we observed that the proportion of registered demands which were satisfied dropped to as low as 0.5 in one test case and to 0.98 in the other test case \cite{journal-arx}.
These findings call for the development of admission control routines capable of achieving better performance with increased reliability. 
Here we define `reliable' to mean that the network can uphold agreements regarding the probability that desired performance levels are achieved. 

In particular, any admission control protocol should reflect the characteristics and performance of the scheduling algorithm employed. 
This has precedent in classical networking, for example in \cite{steenhaut_scheduling_1997}. 
By using the characteristics of the network scheduler to determine the admission control rules, it can be that any demands which are accepted do not disrupt the demands which have previously been accepted for scheduling. To prevent the introduction of significant delays due to classical computation time, we conclude that whichever scheduling algorithm is used, there should exist a fast (polynomial time) feasibility test for admission control, determining whether or not new demands can be admitted.


\section{Fast scheduling}

The computational complexity of the scheduling algorithm used to construct the network schedule fundamentally depends on the number of PGAs which need to be added to the schedule. 
Our architecture relies on the premise that the central controller can compute the network schedule and distribute it to the end nodes in time for it to be executed. With a \textit{slow} or highly complex scheduling algorithm this may not always be possible. 

Indeed, in our evaluation\cite{journal-arx}, we used a scheduler based on \textit{earliest deadline first} scheduling \cite{liu_scheduling_1973}, which has a complexity of $O\big{(}N\log(N)\big{)}$ where $N$ is the number of PGAs to be added to the network schedule. 
We observed that when there were more than 1500 PGAs to add to the network schedule, the estimated time to compute the schedule exceeded the maximum time allocated \cite{journal-arx}. 
This meant that the second admission control rule was limiting the performance of the network, as our implementation of the scheduling algorithm was not capable of producing schedules on time without limiting the number of demands which were accepted by admission control. 
Therefore, it is very apparent that a \textit{faster} scheduling algorithm is needed.
 This would allow the network to successfully schedule more PGAs, and consequently satisfy a greater number of demands.

In our architecture \cite{journal-arx}, the allowable time to compute each network schedule  is the offset, in terms of a number of scheduling intervals, between the start of computation time and the time at which the schedule must be distributed. 
Moreover, each of these network schedules will last for a single scheduling interval. 
It is worth noting that one cannot solve the problem of required schedule computation time by simply increasing the scheduling interval. 
If the total number of PGAs in the schedule is not correspondingly increased, then there will necessarily be idle time introduced into the network schedule. 
On the other hand, if the total number of PGAs is increased, then the total time to compute the schedule will also increase, thereby recreating the problem of insufficient time to compute the network schedule. 
One could mitigate the problem by increasing the time allocated (in terms of number of scheduling intervals) to compute the network schedule.
However, this will increase the response time between submitting a demand and the first time a PGA is scheduled to serve that demand. 
As in classical networks,
the response time is a performance  metric which may be relevant to end nodes.
Therefore increasing it to compensate for a sub-optimal scheduling algorithm is not always a tenable solution. 

We have illustrated the requirement of fast scheduling algorithms in the context of our centrally controlled network architecture. However, the considerations are of even greater importance in a setting where network scheduling is accomplished in a distributed manner. 
This is because distributed implementations may also require increased rounds of classical communication, each of which may be relatively slow.    
Therefore, there will be even less time available to compute a network schedule and hence a faster scheduling algorithm will be required to be able to consistently compute and distribute schedules on time. 


\section{Latency and jitter}
In the architecture we propose in \cite{journal-arx}, the primary objective of the network is to facilitate the generation of at least a minimum number of packets of entanglement, where this minimum is a demand parameter. 
However, we do not regulate the jitter between packet generations or the response time.


Applications in the HA class \cite{HFT, loadBalancing} may impose limitations on the acceptable jitter between the generation of packets of entanglement. 
If typical network traffic includes such applications, then our architecture may require modification to satisfy demands pertaining to these applications.
We identify investigation of the impact of jitter between packet generations and possible mitigation strategies as an open challenge.  

In a similar vein, we do not attempt to optimize the response time.
In our architecture, there will always be a minimum response time of at least one scheduling interval. This is especially notable if the duration of the scheduling interval is on the order of tens of minutes or more. 
For certain applications this may not be desirable, for example applications of QKD such as authenticated monitoring of systems (e.g. \cite{alshowkan_authentication_2022}). 
In these scenarios, there need to be entangled links readily available to ensure continuation of service. An inherent response time before service is provided may prevent such applications from being executed. 
We identify two approaches for modifying our architecture one could take in order to mitigate such inherent response times. 

The first approach may be to make the production of network schedules a more `online' process, where the schedule is updated more frequently. 
Currently, the network schedule is only computed once per scheduling interval. 
This approach would also require development of protocols to allow updating network schedules at the nodes in the network. 
In addition, this approach may require bounding the classical communication and computation required to realize such a protocol. 

The second approach would be to design protocols which implement a functionally pre-loaded network on a GWR type network. 
Recall from Chapter~\ref{ch: arx} that by `pre-loaded' we mean that entangled links are buffered at all times between pairs of end nodes, such that whenever an application requires entangled links there are almost surely enough links available. 
The idea here is that by tailoring submission of demands to the network for entangled link generation, end nodes may be able to engineer a situation in which they almost certainly have sufficiently many entangled links in storage.
The design of protocols implementing this approach is an outstanding challenge. 
A protocol which solves this challenge would potentially open up near term networks to many more applications than would otherwise be possible.

\section{Responsibility of network controllers}
In classical networking, there exists something called the \textit{end-to-end principle} \cite{saltzer_end--end_1984}.
This states that any network should not take proactive measures to ensure that any information is transmitted correctly, and that this is the sole responsibility of high level applications running on the devices which are communicating with each other. 
As the central controller in our architecture does not monitor whether or not end-to-end entangled links are generated in any given PGA, we adhere to this principle. 
Because of this, the responsibility lies squarely with the end nodes to ensure that they generate the same packets as they requested. If an insufficient number of packets are generated to satisfy the demand, the end nodes must submit a new request rather than the central controller taking steps to remedy this itself. 

However, there exists the possibility for end nodes to submit demands with an adaptive rate of packet generation.
When an adaptive rate demand gets accepted by admission control, the central controller sets the PGA rate for the demand such that it will almost surely still be able to generate at least the requested number of packets of entanglement,  regardless of how long the demand may have sat in a queue.
This is somewhat in contravention of the end-to-end principle, as the central controller has the responsibility to ensure that PGAs are scheduled at a sufficiently high rate, rather than the end nodes themselves submitting a demand with a suitable rate. 
In a distributed architecture, it may be possible for end nodes to enact an adaptive approach, for example based on monitoring received schedules in response to demands and based on a rate control algorithm \cite{FlowControlI, gauthier_control_2023}.

In evaluating our architecture, we observed that the demand queue implementation affected the effectiveness of adaptive rate demands. When there were many adaptive rate demands, queuing times increased, which increased required PGA rates and created a cycle where the proportion of satisfied demands decreased with the increase of adaptive rate demands. This was partly due to using a first-in-first-out (FIFO) queue, where admissible demands could be blocked by inadmissible ones. Additionally, we placed no limitations on maximum PGA rates per demand, which could possibly improve the effectiveness of the adaptive rate strategy, by limiting each demand's network impact.

If a network implementation utilises a demand queue, we identify two challenges which need to be addressed. 
The first is designing or selecting a queue implementation and management system which ensures admissible demands can be accepted. 
The second is designing or modifying a protocol to control how demands are adapted to ensure that no single demand can overload the network.


\section{How to implement network capability management?}

One of the key components of our architecture is the existence of the so-called `\textit{Network Capabilities Manager}' (NCM).
This device or process has the responsibility to maintain an accurate picture of the end-to-end entanglement generation capabilities of the quantum network, and to report these to both the central controller and the end nodes, as requested or required. 
However, there is no prior work considering the practical challenges of implementing such a capabilities manager in a quantum network. 
Among the outstanding questions which would need to be answered to implement a network capabilities manager include: What information needs to be exposed by the network components to the NCM? How often does this information need updating? How does the NCM calculate the end-to-end rates and fidelities?

In very near term experimental networks, limited to a small number of devices operated by a single organization in a controlled environment, it is possible to forgo a `smart' NCM and instead simply directly characterize the capabilities of each end-to-end link ahead of running a given experiment. 
However, as the number of paths to characterize grows exponentially with the size of the network, even with only a handful of nodes, direct end-to-end characterization can become infeasible. 
This is especially the case when entanglement generation rates are low, as even characterization of an individual end-to-end path may take a significant period of time. 
Therefore, implementing an NCM which only requires characterizing elementary links over short distances will also greatly improve the ability to operate and run experiments with these near term networks. 

In any quantum network architecture, there will need to exist some level of network capability management to ensure that appropriate routes are selected to generate end-to-end entangled links. 
Furthermore in a distributed system, including best effort architectures, nodes may have to make such decisions informed primarily by knowledge of the  entanglement generation capabilities with their nearest neighbors.
In such a case, it will be necessary to design protocols to ensure that these local decisions lead to a sufficiently high quality end-to-end route.  
Some distributed protocols for routing have been proposed, for example \cite{van_meter_quantum_2022}, and perhaps these could be adapted to achieve our requirements.

\section{How do nodes join the network?}
It is not expected that quantum networks are static over extended periods of time. 
Indeed, we expect that as the technology matures, more and more users will want access to the network, which may require new end nodes to join the network and new internal nodes to be installed. 
Therefore, an outstanding challenge is to design a protocol governing how new nodes join the network. 

When a new node joins the network, it needs to make itself known. 
In our centrally controlled network architecture, this means announcing itself to the network capabilities manager, and in general the central controller. 
In a distributed network, this requires first establishing who its neighbors are, and then announcing itself to them. 
As part of this announcement, any node will need to expose some information about its hardware and capabilities in order for routing decisions to be made involving that node. 
However, it is an outstanding question as to exactly what information needs to be exposed at this point. 

Furthermore, a certain level of clock synchronization is required to ensure that key events happen at the correct time. 
For instance, at the physical layer when individual attempts to generate an elementary entangled link are being performed, the two nodes involved may need to have at least nanosecond levels of synchronization between themselves. 
The clock precision required to execute the schedule is not as high as that at the physical layer, however it may still be the case that microsecond precision is required. 
Fundamentally, however, the timing precision available to the upper layers of the network nodes will determine the shortest amount of time that can be scheduled for any path involving that node.
Therefore, a lack of timing precision at the upper layers may become limiting if this minimum time becomes much longer than the computed required time for entangled link generation, thereby limiting the performance which can be extracted from the network. 

If the network is producing and distributing a network schedule, then there also needs to be a method of establishing a common epoch or absolute clock across all of the nodes to ensure that the schedule is carried out correctly. 
Without synchronization, nodes may execute schedules correctly according to their local clocks, but misalignment can severely reduce or prevent entanglement generation.
Furthermore, sufficient clock misalignment may lead to creating entanglement between unexpected parties, which may become a security issue and introduce an attack vector for otherwise secure applications. 
However, it is unclear exactly how to ensure that this common epoch is established and enforced.
Therefore, it is an open challenge to design such a protocol. 

\section{When is a service agreement made between the network and the end nodes?}
Based on the view that the aim of a quantum network is to provide the service of end-to-end entanglement generation to end nodes, we propose that the overriding performance criterion is the satisfaction of demands. 
One question that still needs to be addressed is: from which point does the network have a responsibility to provide a pair of nodes with service.
In the evaluation we performed, we chose to measure the proportion of \textit{registered} demands which generated at least the requested number of packets of entanglement (recall from Chapter~\ref{ch: arx} we call this \textit{obtaining minimal service} from the network).
This gives rise to a notional service agreement of ``\textit{given your demand has been registered, you will be almost certainly be able to obtain minimal service}''.
However, this may not be the correct service agreement to make. 
For instance, as we observed in our evaluations, the likelihood of being able to meet this service agreement depends strongly on the volume of demands currently placed on the network.

An alternative to this would be to consider only making service agreements after admission control has occurred. 
This increases the feasibility of meeting service agreements, since admission control can account for what can realistically be scheduled.
However, this comes at the cost of neglecting the demands which are registered but never make it through admission control. 
Depending on the envisioned applications in a deployment, this may or may not be acceptable. 
Therefore, it is an outstanding question to determine what the correct point is to make the service agreement, and what form the agreement should take. 


\chapter{Conclusions}
\label{ch: conclusions}
In this paper we have leaned on our experiences in implementing a quantum network architecture to identify outstanding challenges in the domain of designing and implementing quantum network architectures. 
We have laid out some of the most pressing challenges and outstanding questions, and provided possible directions towards solutions.  
It is our hope that the ideas presented here can be used to stimulate further research towards solving these problems, developing new protocols, and to also act as a catalogue of challenges requiring solutions.

\chapter{Acknowledgments}
This work was supported by the Quantum Internet Alliance (QIA). QIA has received funding from the European Union’s Horizon Europe research and innovation programme under grant agreement No. 101102140. SW also acknowledges funding from the NWO VICI grant.


    

\bibliographystyle{IEEEtran}
\bibliography{references}
\end{document}